\documentclass[prl,aps,twocolumn]{revtex4}

\def\<{\langle} \def\>{\rangle}

\usepackage{graphics}

\begin{document}

\title{Secure communication using coherent states}

\author{Geraldo A. Barbosa, Eric Corndorf, Prem Kumar, and Horace P. Yuen}
\affiliation{Center for Photonic Communication and Computing, ECE Department, Northwestern University, Evanston, IL 60208-3118 \\
E-mail: barbosa@ece.northwestern.edu}
\author{G. Mauro D'Ariano,  Matteo G. A. Paris, and Paolo Perinotti}
\affiliation{Quantum Optics $\&$ Information Group, \\
INFM, Universit\`a di Pavia, via Bassi 6, I-27100 Pavia, Italy}


\begin{abstract}
We demonstrate that secure communication using coherent
states is possible. The optimal eavesdropping strategy for an
$M$-ry ciphering scheme shows that the  minimum probability of
error in a measurement for bit determination can be made
arbitrarily close to the pure guessing value $P_e=1/2$. 
This ciphering scheme can be optically amplified without degrading
the security level. New avenues are open to secure communications
at high speeds in fiber-optic or free-space channels.
\end{abstract}

\maketitle
Secure communication protocols protected by physical laws instead
of mathematical complexities, such as the BB84 quantum protocol
for key distribution,\cite{bennett-brassard}  seem to have
encountered a bottle-neck that hampers their utilization in real
networks. The same no-cloning theorem that guarantees security
forbids signal amplification necessary in long-haul communication
links. No alternate quantum scheme using quantum repeaters or
entangled states, having practical applicability within a
reasonable time span, has been envisaged. The question ``is it
possible to create a system with available technology that could
provide unconditional security  in long distance communication?''
has a positive answer. It relies on the use of quantum noise
inherent in the coherent states of light, as demonstrated in this
paper. 

In the presence of noise, it has been shown information
theoretically\cite{wyner} that new shared secret keys can be
created between two users. A specific protocol has been proposed
by Yuen,\cite{yuen-kim,yuen1} which has been called the YK
protocol. A particular implementation of this YK protocol has been
demonstrated,\cite{tomita-hirota} in which detector noise is
utilized and hence it is not unconditionally secure. However, the
YK protocol can be made unconditionally secure by utilizing the
fundamental, unavoidable quantum noise in a quantum signaling
scheme.\cite{yuen2} The other known secure scheme  is, of course,
the BB84,\cite{bennett-brassard} or its variant  that involves
features of B92,\cite{bennett} where coherent states instead of
Fock states are employed. The implementations of the YK
protocol,\cite{tomita-hirota} and BB84, both suffer from the
intrinsic limitation that very weak signals with no more than one
photon per mode have to be used, making them severely rate-limited
in a lossy channel. This problem can be alleviated in a new
protocol, where mesoscopic coherent states are employed to
overcome loss and to allow ordinary amplification, switching and
routing. In addition to the use of the fundamentally-unavoidable
quantum noise in coherent states, a crucial new ingredient in this
protocol is the explicit use of a shared secret key for the
cryptographic objective of key expansion. A shared secret key is
also needed in the BB84 and the YK protocols for the purpose of
user and message authentication amidst the protocol execution.

In the ciphering scheme for this new protocol
the sender (Alice) uses an explicit secret key (a short key ${\bf
K}$, appropriately expanded into a longer key ${\bf K}^\prime$ by
use of another encryption mechanism such as a stream cipher) to
modulate the parameters of, in general, a multimode coherent
state. Coherent states span an infinite dimensional Hilbert space
which we refer to as a {\em qumode}. A qumode can be associated
with any physical property of light such as polarization, phase,
frequency or time. For the free-space implementation to be
presented, the qumodes are the two orthogonal modes of
polarization. In this case, Alice uses the running key ${\bf
K}^\prime$ to specify a polarization basis from a set of $M$
uniformly spaced two-mode bases spanning a great circle on the
Poincar\'e sphere.
Each basis consists of a polarization state and its antipodal
state at an angle $\pi$ from it, representing the 0 and 1 bit
value for that basis. The message ${\bf X}$ is encoded as ${\bf
Y}_{K^\prime}({\bf X})$. This mapping of the stream of bits onto
points of the Poincar\'e sphere is the {\em key} to be shared by
Alice and the receiver (Bob).  Because of his knowledge of ${\bf
K}^\prime$, Bob is able to make a precise demodulation operation
producing the plaintext ${\bf X}$. Bob applies ${\bf K}^\prime$ to
the received sequence of arbitrary polarization states to return
them to the linearly orthogonally polarized condition,
representing the two original bits of the message ${\bf X}$.

We present analysis  that covers both polarization as well as
phase modulation of optical signals. In the case of {\em
polarization} states, information is encrypted and encoded on two
orthogonal polarization modes of radiation with annihilation
operators $a_1$ and $a_2$.  A coherent state $|\psi_0\>$ with
amplitude $\alpha$ in mode $a_2$
($|\psi_0\>=|0\>\otimes|\alpha\>$) is ``rotated'' by a unitary
transformation $U_{\varphi_b}$ ($\varphi_b=0$ or $\pi$) to create
the bit of a message. This rotation is performed by, e.g.,
$U_{\varphi_b}=\exp[(\varphi_b/2)(a_1^\dag a_2 - a_1 a_2^\dag)]$,
giving
\begin{equation}
|\psi_b\>=U_{\varphi_b}|0 \>\otimes|\alpha\> =
|\alpha\sin\frac{\varphi_b}{2} \>\otimes
|\alpha\cos\frac{\varphi_b}{2}\>.
\end{equation}
For {\em phase} encoding within a given polarization state, one
could start by splitting a coherent state $|\alpha\>$ into  a
two-mode coherent state $|\Psi_0 \>=|\alpha/\sqrt{2}\>_1 \otimes
|\alpha/\sqrt{2}\>_2$. Bit encoding can be represented by the
operation
\begin{equation}
\label{Psi_b} |\Psi_b \>=e^{- i J_z \varphi_b}|\Psi_0 \>=|e^{- i
\varphi_b/2}\alpha/\sqrt{2}\>_1 \otimes |e^{ i
\varphi_b/2}\alpha/\sqrt{2}\>_2,
\end{equation}
where $J_z=( a_1^{\dagger} a_1 -a_2^{\dagger} a_2 )/2$.
Demonstration of the security of the ciphering scheme in both
kinds of  physical encoding, polarization or phase,  can be
treated with the same formalism and leads to the same result.

Let us analyze the phase ciphering, applied by the same modulator
generating the bit sequence. The ciphering angle $\phi_{\nu}$
could have  $\nu$ as a discrete or a continuous variable
determined by some general distribution. A ciphered two-mode state
is
$|\Psi_{b\nu} \>=e^{- i J_z (\varphi_b+\phi_\nu) }|\Psi_0 \>$ and
the corresponding density operator for all possible choices of
$\nu$ is $\rho_b$.
The problem is to find the minimum probability of error $P_e^E$
that an eavesdropper (Eve) can achieve in bit determination,
given that ciphered states are used.

No restriction is imposed on the physical devices available to
Eve, including perfect detectors and unlimited computational
power. A close-to-source attack is considered, where no losses
have yet occurred that normally would during propagation of the
signal. These are the ideal conditions for an eavesdropper.

\begin{figure}
\centerline{\scalebox{0.4}{\includegraphics{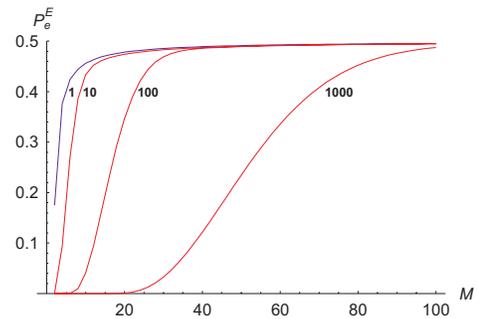}}}
\caption{\label{fig1} $P_e^E$ as a function of $M$ for $|\alpha|^2
=1,\: 10,\:100,\:100$.} \vspace*{-0.125in}
\end{figure}

The optimal POVM for discriminating between $\rho_0$ and $\rho_1$
is given by Helstrom's binary discrimination
procedure\cite{helstrom} applied to $\Delta\rho=\rho_1-\rho_0$.
Calling $\Pi_1$ and $\Pi_0$ ($\Pi_1+\Pi_0 ={\rm I}$) the
projectors over eigenstates with the positive and negative
eigenvalues of $\Delta \rho$, the probability of error $P_e^E$ is
\begin{equation}
P_e^E={\rm Tr}\left[p_1 \Pi_0\rho_1+p_0 \Pi_1\rho_0\right],
\end{equation}
where $p_1$ and $p_0$ are {\em a-priori} probabilities to find a
state in $\rho_1$ or $\rho_0$, respectively. In one of Yuen's
scheme,\cite{yuen} closest values of a given $k$ are associated
with distinct bits from the bit at position $k$. For example,
$\phi_{k}= \pi  \left[k/M + (1/2) \left(  1-(-1)^k \right)
\right], \: k=0,1, ...,M-1$. In this encoding protocol, two
closest neighboring states represent distinct bits.
Figure~\ref{fig1} shows the minimum probability of error as a
function of the number of ciphering levels $M$. $P_e^E$ goes very
fast to the asymptotic pure-guessing limit of $1/2$ as $M$
increases. The quantum noise present in the coherent state of
light is what turns these states indistinguishable to an
eavesdropper.

Bob, on the other hand, by knowing the key has a more complete
information on the state of light sent and can extract the
information with great precision. His probability of error is
\begin{equation}
P_{e}^{B}={1 \over 2}\left(1-\sqrt{1-e^{-2 |\alpha|^2}}\right).
\end{equation}
For sufficiently large values of $\alpha$ the  minimum probability
of error $P_e^{B}$ is negligible, leading to an excellent signal
recovery by Bob.

Note that the present analysis does not include Eve's attacks
using trial keys, which would be exponentially complex. Also, the
stream cipher output in the system is not open to observation,
thus its seed key is not open to the usual known plaintext
attacks. The complete security analysis for the most general
attacks would be presented elsewhere.

This scheme opens up the path for long-distance secure
communication based on protection by physical laws instead of
mathematical complexities.

\section*{Acknowledgments}
This work was supported by DARPA/AFRL under grant \#
F30602-01-2-0528.


\end{document}